\documentclass[nofootinbib,prd,
]{revtex4}

\usepackage{mathrsfs}
\usepackage{tipa}
\usepackage{bm}
\usepackage{amsmath}
\usepackage{amssymb}
\usepackage{amsthm}
\usepackage{hyperref}
\usepackage{amsfonts}
\usepackage{mathtools}
\usepackage{latexsym}
\usepackage{relsize}
\usepackage{booktabs}
\usepackage{tikz}
\usepackage{wrapfig}

\begin{document}

\title{Cosmological production of black holes: \\a way to constrain alternative theories of gravity}

\author{Konstantinos F. Dialektopoulos}
\email{dialektopoulos@na.infn.it}
\affiliation{Dipartimento di Fisica, Universit\'a di Napoli {}``Federico II'', Compl. Univ. di Monte S. Angelo, Edificio G, Via Cinthia, I-80126, Napoli, Italy,}
\affiliation{INFN Sezione  di Napoli, Compl. Univ. di Monte S. Angelo, Edificio G, Via Cinthia, I-80126, Napoli, Italy.}
%\affiliation{Lepage Research Institute, 17. Novembra 1, 08116 Pre\v sov, Slovakia}

\author{Antonios Nathanail}
\email{nathanail@th.physik.uni-frankfurt.de}
\affiliation{Institut f\" ur Theoretische Physik, Goethe Universit\" at Frankfurt, Max-von-Laue-Str.1, 60438 Frankfurt am Main, Germany}

\author{Athanasios G. Tzikas}
\email{tzikas@fias.uni-frankfurt.de}
\affiliation{Frankfurt Institute for Advanced Studies (FIAS), Ruth-Moufang-Str. 1, 60438 Frankfurt am Main, Germany}
\affiliation{Institut f\" ur Theoretische Physik, Goethe Universit\" at Frankfurt, Max-von-Laue-Str.1, 60438 Frankfurt am Main, Germany}

\date{\today}

\begin{abstract}
Primordial black holes are considered to be pair created quantum-mechanically during inflation. In the context of General Relativity (GR), it has been shown that the pair creation rate is exponentially decreasing during inflation. Specifically, tiny black holes are favored in the early universe, but they can grow with the horizon scale, as inflation approaches its end. At the same time, cosmological, and not only, shortcomings of GR have triggered the pursuit for a new, alternative theory of gravity. In this paper, by using probability amplitudes from the No Boundary Proposal (NBP), we argue that any alternative gravity should have a black hole creation rate similar to that of GR; that is, in the early universe the creation of small black holes is in favor, while in the late universe larger black holes are being exponentially suppressed. As an example, we apply this argument in $f(R)$-theories of gravity and derive a general formula for the rate in any $f(R)$-theory with constant curvature. Finally, we consider well known $f(R)$-models and using this formula we put constraints on their free parameters.
\end{abstract}

\maketitle

\section{Introduction}
\label{intro}

According to the concordance model in cosmology, $\Lambda$CDM, the universe is endowed with a positive cosmological constant, which has a current value of $\Lambda \simeq 1.11 \times 10^{-52}m^{-2} \simeq 2.9 \times 10^{-122}$ in reduced Planck units \cite{Ade:2015xua}. However, it is expected that, in the beginning of the universe, and especially during inflation, it started out large enough and since then decreases, until its current value. In addition, the models of inflation predict cosmic density perturbations, as well as quantum fluctuations of the field, responsible for the inflation, which lead to topological changes of the spacetime and these respectively to the creation of primordial black hole pairs. 

Primordial black holes (PBHs) are theoretical objects with masses smaller than the solar mass  down to the Planck mass. These objects are unlikely to form from the gravitational collapse of a star today, since low-mass black holes can form only if matter is compressed to enormously high densities by very large external pressures \cite{CaH74}. Such conditions of high temperatures and pressures can be found in the early stages of a violent universe and, thus, it is believed that PBHs may have been produced plentiful back then. Proposals for their formation have been addressed over the years, such as the collapse of primordial inhomogeneities \cite{Car04}, cosmological phase transitions \cite{RSK01,Khl10} and close domain walls \cite{KRS05}.

Apart from the procedure of the spontaneous formation of black hole pairs, there should exist a force which would pull them apart, in order not to fall back together and annihilate. In our scenario, we assume that the cosmological constant plays the role of this force, and because of the rapid cosmological expansion in the early universe, the created black hole pair remains separate.  

During the nineties, the study of black hole pair creation in different backgrounds was of much interest \cite{Mann:1995vb,Dowker:1993bt,Bousso:1996au,Garfinkle:1993xk,Dowker:1994up,Hawking:1994ii,Hawking:1995ap,Caldwell:1996pt,MaN11}. Important results have been obtained  considering creation of different types of black holes in various backgrounds. In particular, one has to consider two different spacetimes; one for the inflationary background (e.g. de Sitter) and another one with  black holes in this background (e.g. Schwarzschild-de Sitter (SdS)). For each one of these two spacetimes, we have to construct an instanton\footnote{Instantons are complex solutions of the Euclidean Einstein equations and, thus, their signature is $(++++)$. They appear in the path integral approach as leading corrections to the classical behavior of the system and they are commonly used to describe non-perturbative gravitational effects.} and its analytical continuation to the Lorentzian solution, if possible. The Hartle-Hawking No Boundary Proposal \cite{HarHawk83} suggests that in the semi-classical approximation\footnote{Semi-classical in the sense that we treat the geometry of  spacetime classically, while the rest of the fields quantum mechanically.}, each universe can be described by a wave function $\Psi$ and the probability amplitude for the creation of this universe is given by 
$P = |\Psi|^2 = e^{-2I^{\text{Re}}}\,,$ where ``Re" denotes the real part of the instanton-action $I\,$. In this way, the pair creation rate, $\Gamma$, is calculated as the ratio of the two probabilities; the probability of the universe containing a pair of black holes, described by the instanton-action $I_{\mathrm{obj}}\,$, and the probability of the background universe, described by $I_{\mathrm{bg}}\,$,
\begin{equation} \label{rateformula}
\Gamma = \frac{P_{\mathrm{obj}}}{P_{\mathrm{bg}}} = \exp \left[ -2 \left(I_{\mathrm{obj}}-I_{\mathrm{bg}} \right)\right]\,.
\end{equation} 

\begin{figure}[!ht]
\includegraphics[scale=0.44]{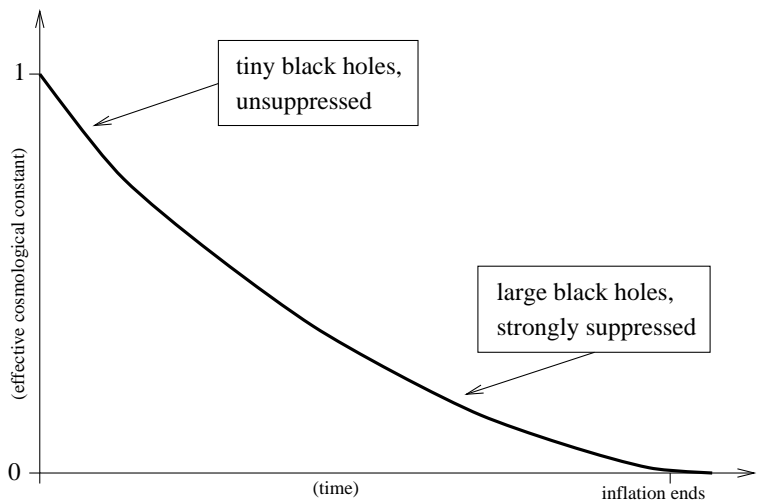}
\caption{The diagram is taken from \cite{Bousso:1996wz} and quoting the authors: (It depicts) \textit{``The classical evolution of the effective cosmological constant in a typical model of chaotic inflation. We have indicated qualitatively how the nucleation size and pair creation rate of black holes depend on the effective cosmological constant."}}
\label{fig1}
\end{figure}

In the framework of GR, Bousso and Hawking \cite{BouHa95,Bousso:1996au} found that the pair creation rate of two neutral (SdS) black holes in an inflationary (de Sitter) background is $\Gamma = e^{-\pi/\Lambda}\,$. Even though $\Lambda$ is varying, its decrease is considered to be very slow in time and, thus, in the calculations it was considered as constant. The interpretation of this result is that in the early universe, where $\Lambda$ is considered to be of order unity in Planck units, the creation of tiny black holes, $r_{\mathrm{bh}} = 1/\sqrt{\Lambda}\,$, is favored. As inflation reaches its end, the size of the black holes that would be created, increases significantly, while the probability rate becomes vastly suppressed. This can be seen in the Fig. \ref{fig1}.

The last two decades though, much attention has been given to modified theories of gravity. The inability of Einstein's 	General Relativity (GR) to give convincing explanations to recent cosmological observations (value of the cosmological constant, nature of the dark matter, coincidence problem, missing satellites, etc.), together with the fact that it cannot lead to a convincing quantum theory for the gravitational interactions, initiated the pursuit of a new theory of gravity. The interesting reader is advised to check the exhausting reviews \cite{Capozziello:2011et,Clifton:2011jh} as well as references therein.

In this work, we propose a way to discriminate between these attempts: the pair creation rate. \textit{Any viable alternative theory of gravity should predict a decreasing pair creation rate, in order to be consistent with observations.} In this article, we deal with the so-called $f(R)$-theories, as an example to our proposal. One of the most straightforward ways to extend GR is to relax the assumption that the gravitational action must be linear in the Ricci scalar, i.e. the Einstein-Hilbert action. Indeed, if we substitute the Ricci scalar, $R$, by an arbitrary function $f(R)$, we obtain a more general class of theories, which lead to richer phenomenology \cite{Capozziello:2009nq,Nojiri:2006ri,Nojiri:2010wj} compared to GR. It is not the purpose of the paper to get into the details of the theory; in the references given, there is big amount of information regarding not only this specific class of modifications, but also other attempts to modify GR. Nevertheless, the choice of $f(R)$ gravity is not arbitrary. Its simplicity, compared to other higher-order theories, as well as its ability to give interesting results the last few years, made it an interesting alternative to GR. However, it is still nothing but a toy-theory; a tool to understand the limits and the underlying principles of modified gravity.

We follow the same procedure, in order to calculate the pair creation rate of SdS black holes in a de Sitter background in the context of $f(R)-$gravity theories with constant curvature. As we shall see, and as expected, the rate depends on the functional form of $f(R)$. The free parameters of each theory are embodied in the pair creation rate. The purpose of this paper is to conservatively set constraints on the free parameters of $f(R)-$models, by imposing that the production rate gives physically acceptable results, in agreement with GR. However, the fact that we consider $R =\text{constant}$ makes our results not generic. Specifically, this assumption kills all the higher derivative terms in the equations of motion and thus the results obtained (here, black hole solutions) are exactly the same as in GR. The example though is good enough in order to apply our criterion at a toy-model. We do not claim that our results are generic, neither that the $R = \text{constant}$ assumption is realistic during the inflationary period. However, at first approximation, it gives some interesting results and it is exactly the same as  assuming that in GR $\Lambda _{\text{eff}} = \text{constant}$ \cite{Mann:1995vb,Bousso:1996au,BouHa95}; at least at the solutions level.

This article is organized as follows: in the section \ref{BHsinfR} we introduce the basics of the class of $f(R)-$theories; action, equations of motion and already known constraints. In addition, we review some already known black holes solutions in those theories with constant curvature. As we will see, the form of both SdS and de Sitter solutions remain the same as in GR, with the only difference being that the gravitational coupling and the cosmological constant acquire a contribution from the new degree of freedom of the theory. In the next section, \ref{InstMeth}, we briefly describe the instanton method, i.e. how to find instantons in generic spacetimes. We also give two examples, regarding the instanton of de Sitter and SdS spacetime. Finally, in section \ref{fRpcrate} we apply the same method in $f(R)-$gravity and obtain the pair creation rate of two neutral black holes in an expanding background. We specifically use the physical significance of the computed rates to set constraints on the free parameters of well-known cosmological $f(R)-$models. In section \ref{conclus} we conclude and discuss future perspectives.

\section{Black Holes in $f(R)$ gravity}
\label{BHsinfR}

As already mentioned in the introduction, if we substitute in the Einstein-Hilbert (EH) action the Ricci scalar, $R$, with an arbitrary function $f$ of it, we obtain the action of the $f(R)-$theories,
\begin{equation}\label{f(R)action}
\mathcal{S} = \int d^4x \sqrt{-g} \left(\frac{1}{16 \pi G_{\text{N}}} f(R) + \mathcal{L}_{\mathrm{m}} \right)\,,
\end{equation}
where $G_{\text{N}}$ is  Newton's constant $(c = 1,\,\hbar =1,\,k_{\mathrm{B}}=1 )\,$, $g$ is the determinant of the metric and $\mathcal{L}_{\mathrm{m}}$ is the Lagrangian density of all the matter fields. By varying this action with respect to the dynamical field, i.e. the metric, one gets
\begin{equation}\label{f(R)eom}
R_{\mu\nu}f'(R) -\frac{1}{2}g_{\mu\nu}f(R) - \left(\nabla_{\mu}\nabla_{\nu}-g_{\mu\nu}\square \right) f'(R) = 8 \pi G_{\text{N}} T_{\mu\nu}^{\mathrm{M}}\,,
\end{equation}
where the prime denotes differentiation with respect to the argument, i.e. $f'(R) = df(R)/dR$. The tensor $T_{\mu\nu}^{\mathrm{M}}$ is the energy-momentum tensor of all the matter fields and is defined as  $T_{\mu\nu}^{\mathrm{M}} = - \frac{2}{\sqrt{-g}}\frac{\delta \left(\sqrt{-g}\mathcal{L}_{\mathrm{m}}\right)}{\delta g^{\mu\nu}}\,$. One more interesting equation is the trace of the equations \eqref{f(R)eom}, which reads
\begin{equation}\label{f(R)trace}
R f'(R) - 2 f(R) + 3 \square f'(R) = 8 \pi G_{\text{N}} T^{\mathrm{M}} \,,
\end{equation}
and is the dynamical equation of the Ricci scalar; the new degree of freedom of the theory.

If we rewrite the equations in an Einstein-like form, i.e.
\begin{equation}
G_{\mu\nu} = \frac{8 \pi G_{\text{N}} }{f'(R)}T_{\mu\nu}^{\mathrm{M}} + \frac{1}{f'(R)} \left[\frac{1}{2}g_{\mu\nu}\left(f(R)-  Rf'(R)\right) + \left( \nabla_{\mu}\nabla_{\nu} - g_{\mu\nu}\square\right) f'(R) \right] = \frac{1}{f'(R)}\left[ 8 \pi G_{\text{N}} T_{\mu\nu}^{\mathrm{M}} + T_{\mu\nu}^{{\mathrm{curv}}}\right]\,,
\end{equation}
we immediately see that, we have a purely geometrical contribution to the effective energy momentum tensor, coming from the new scalar degree of freedom; thus, we will have much richer phenomenology compared to GR.

In this paper, we will only consider theories with constant curvature, i.e. $R = R_{{\mathrm{dS}}}\,$, where '${\mathrm{dS}}$' stands for de Sitter. Specifically, $R_{{\mathrm{dS}}}$ is a solution of the trace Eq. \eqref{f(R)trace}, which for constant curvature becomes
\begin{equation}\label{R=R0trace}
Rf'(R)- 2f(R) = 0\,.
\end{equation}
These solutions are called de Sitter points \cite{DeFelice:2010aj}. 
The equations of motion \eqref{f(R)eom} in vacuum, i.e. $T_{\mu\nu}^{\mathrm{M}}=0\,$, become
\begin{equation}
R_{\mu\nu} = \frac{R_{{\mathrm{dS}}}}{4}g_{\mu\nu}\,,
\end{equation}
which means that as soon as $R_{{\mathrm{dS}}} \neq 0 $, these theories accept solutions of the (Anti-)de Sitter type, or Schwarzschild-(Anti-)de Sitter type, or Kerr-(Anti-)de Sitter type, depending on the sign of $R_{{\mathrm{dS}}}$ and, thus, they introduce an effective cosmological constant 
\begin{equation}\label{lambdaeff}
\Lambda_{\text{eff}} = \frac{R_{{\mathrm{dS}}}}{4} = \frac{f(R_{{\mathrm{dS}}})}{2 f'(R_{{\mathrm{dS}}})}\,,
\end{equation} 
where we used Eq. \eqref{R=R0trace}.

This kind of theories has been studied extensively in the literature in many different contexts. Some reviews on the topic are \cite{Clifton:2011jh,Sotiriou:2008rp,DeFelice:2010aj,Capozziello:2011et,Capozziello:2009nq}. During the years, and in order to constitute viable alternative to cosmological models, there have been found some constraints on the functional form of $f(R)$. Two of them, that we may find useful later, are the following: 
\begin{itemize}
\item
It has to be $f'(R)>0$, otherwise the theory will contain ghosts.
\item
As we have already mentioned, the $f(R)-$theories contain an extra scalar degree of freedom, compared to GR; the Ricci scalar or scalaron. It turns out that this scalar field is massive and its mass is given by 
\begin{equation}
\mathcal{M}^2 = \frac{1}{3}\left(\frac{f'(R_{\mathrm{dS}})}{f''(R_{{\mathrm{dS}}})}-R_{{\mathrm{dS}}} \right)\,.
\end{equation}
In order to avoid tachyonic instabilities, i.e. $\mathcal{M}^2<0$, we have to assume that $f''(R)>0$.
\end{itemize}
The interested reader should check the literature for more constraints.

Let us now proceed by quantitatively finding some black hole solutions \cite{delaCruzDombriz:2009et}. Consider a static and spherically symmetric metric of the form
\begin{equation} \label{metric}
ds^2 = - A(r) dt^2 +  B^{-1}(r) dr^2 + r^2 d \Omega^2\,,
\end{equation}
where $A(r), \,B(r)$ are two arbitrary metric potentials to be specified and $d\Omega^2 = d\theta^2 + \sin^2\theta d\phi^2$ is the metric of a 2-sphere.  The 00- and 11-components of \eqref{f(R)eom} are
\begin{align}\label{00}
f'\left(R_{\mathrm{dS}}\right) \left(\frac{A' B'}{4}+B \left(\frac{A''}{2}+\frac{A'}{r}-\frac{A'^2}{4 A}\right)\right)+\frac{1}{2} A f\left(R_{\mathrm{dS}}\right) &= 0 \, \\ \label{11}
f'\left(R_{\mathrm{dS}}\right) \frac{ r B \left(A'^2-2 A A''\right)-A B' \left(r A'+4 A\right)}{r A^2 B}-\frac{2 f\left(R_{\mathrm{dS}}\right)}{4 B} &= 0
\end{align}
and by taking a linear combination of these, we find that
\begin{equation}
B(r) = c_1 A(r)\,,
\end{equation}
with $c_1$ being a constant of integration. For simplicity we can choose $c_1 = 1$, in order for the metric coefficients to be $g_{rr} = 1/g_{tt}$ and thus recover Minkowski in the flat limit. By substituting this back in one of the Eq. \eqref{00} or Eq. \eqref{11} we get
\begin{equation}\label{A(r)}
A(r) = c_2 - \frac{c_3}{r} - \frac{f(R_{\mathrm{dS}})}{6 f'(R_{\mathrm{dS}})}r^2= c_2- \frac{c_3}{r} - \frac{R_{\mathrm{dS}}}{12}r^2 = c_2 - \frac{c_3}{r}-\frac{\Lambda _{\text{eff}}}{3}r^2\,,
\end{equation}
where $c_2,\, c_3$ are constants and we used Eq. \eqref{lambdaeff}. Without loss of generality, we can fix these constants to $c_2 = 1$ and $c_3 = 2 G_{\text{eff}} M\,$, where $G_{\text{eff}}$ is the effective gravitational coupling. 

Summarizing, we showed that $f(R)-$gravities with constant curvature have Schwarzschild-de Sitter solutions of the form
\begin{equation}\label{f(R)SdS}
A(r) = 1 - \frac{2 G_{\text{eff}}M}{r} - \frac{\Lambda _{\text{eff}}}{3}r^2\,,
\end{equation}
with $G_{\text{eff}} >0 $ being the effective gravitational coupling and $\Lambda _{\text{eff}}>0$ the effective cosmological constant, as well as de Sitter solutions in the case where $M \rightarrow 0\,$. What we will do now is to use these black hole solutions, in order to calculate the pair creation rate in the context of $f(R)-$theories.

Before we proceed, let us make a comment on the significance of the above solutions. Someone may claim that, since in generic $f(R)-$theories the Birkhoff's theorem does not hold, then these results are not correct. However, this is not exactly true; for the theories with constant curvature, which accept SdS-like solutions, the Birkhoff's theorem holds \cite{Capozziello:2011wg,Nzioki:2013lca} and as we already mentioned, we are dealing only with these theories. Of course, it would be more general to consider cosmological spacetimes, which are solutions for all kind of alternative theories, but as a first encounter with the problem, we think this is a decent result. Once again, we do not claim that our results hold for any $f(R)-$theory, but only for those with constant curvature, who accept static solutions.

\section{The Instanton Method}
\label{InstMeth}

Let us briefly resume the procedure of constructing instantons and thus calculating creation rates. Gravity appears many difficulties in the process of pure quantization through an effective field theory, due to the dimensionality of $G_{\text{N}}$ in $4D$, which makes the theory non-renormalisable \cite{Kiefer13}. However, the semi-classical approximation of quantum gravity, known also as Euclidean quantum gravity \cite{GibHawk13}, is considered to be the best semi-classical limit of the future theory of quantum gravity. According to this approach, any universe can be seen as a quantum system with a wave function $\Psi$, which is given by a path integral over all possible, positive-definite Euclidean metric configurations \cite{HarHawk83}
\begin{equation} \label{Int_psi}
\Psi=\int \mathcal{D}[g_{\mu\nu}] \ e^{-\mathcal{S}_{\mathrm{E}} [g_{\mu\nu}]} \approx e^{-I} \,,
\end{equation}
where $\mathcal{S}_{\mathrm{E}}$ is the Euclidean version of the gravitational action and $I$ is the instanton-action. In most cases the above integral cannot be easily evaluated and, thus, we search for saddle points known as gravitational instantons,  that extremize the action $\mathcal{S}_{\mathrm{E}}\,$. An instanton can be seen as a classical path, interpolating between two states and, thus, providing us with the dominant contribution in the wave function. Therefore,  we can approximate the path integral with the exponent of the instanton-action.

In what follows, we give the necessary steps to construct a gravitational instanton for a spherical metric of the form \eqref{metric}, with $B(r) = A(r)\,$:

\noindent \textit{1.} Firstly, we Wick rotate the time axis, $\tau \rightarrow it\,$, in order to analytically continue from Lorentzian to Euclidean section. In this way, we obtain a Euclidean signature $(++++)\,$. The Euclidean version of the modified action \eqref{f(R)action} will be
\begin{equation} \label{Iaction}
I = - \frac{1}{16 \pi G_{\text{N}}} \int d^4x \sqrt{g} \  f(R)\,.
\end{equation}

\noindent \textit{2.} Secondly, we find all the possible conical singularities by taking the metric potential to be zero, $A(r_{\mathrm{h}}) = 0\,$. These singularities signal the existence of apparent horizons and can be eliminated by demanding $\tau$ to be periodic with a period $\beta$, which is given by the inverse of the Hawking temperature, i.e. $\beta= \frac{2 \pi}{\kappa}= T^{-1}$. The surface gravity $\kappa$ of the horizon is given by $\kappa=\frac{1}{2} \left| \frac{dA(r)}{dr} \right|_{r=r_{\mathrm{h}}} $.

\noindent \textit{3.} Subsequently, we restrict ourselves at regions where the metric is positive-definite by taking the condition $A(r)>0\,$. At these regions the signature of the metric is fixed and so we can produce physical results.

\noindent \textit{4.} Finally, we analytically continue backwards by matching the instanton ($\equiv$Euclidean solution) to the Lorentzian solution at $t={\mathrm{const.}}$ hypersurface.

In the forthcoming section, we will deal only with compact spacetimes that have  $S^3$ (de Sitter like)  and $S^1 \times S^2$ (Schwarzschild-de Sitter like) spatial topology. The Euclidean solution for such spacetimes has been  described in \cite{BouHa95}. Here we will give a quick overview of the characteristics of a de Sitter and a Schwarzschild-de Sitter universe, along with the appropriate coordinate transformations that will help us see the regularity of their instantons.
\par Starting with de Sitter spacetime, the form of $A(r)$ in Eq. \eqref{metric} is 
\begin{equation}\label{AdS}
A(r)=1-\frac{\Lambda}{3} r^2\,,
\end{equation}
with cosmological horizons spotted at $r_{\mathrm{h}}=\sqrt{3/\Lambda}\,$. To eliminate the singularity we make time $\tau$ periodical with period $\beta=2\pi \sqrt{3/\Lambda}\,$. We can apply a coordinate transformation by using the periodic variables 
\begin{equation}
\tau = \sqrt{\frac{3}{\Lambda}} \ \psi \, ,\quad r = \sqrt{\frac{3}{\Lambda}} \cos\xi \,,
\end{equation} 
with periods $2\pi$ and $\pi/2$ respectively, that will allow us to see better the regularity of this instanton. Now Eq. \eqref{AdS} reads $A(r) = \sin ^2 \xi$ and de Sitter instanton becomes
\begin{equation}\label{dSinstanton}
ds^2 = \frac{3}{\Lambda} \left( d\xi^2 + \sin^2\xi \ d\psi^2 + \cos^2\xi \ d\Omega^2 \right) \,,
\end{equation}
where \ $d\Omega^2 = d\theta^2 + \sin^2\theta d\phi ^2$. The apparent singularity at $\xi=0$ ($r=\sqrt{3/\Lambda}$) can be seen as an axis of spherical polar coordinates $(\xi,\psi)$ and the manifold is perfectly regular there.

Last but not least, we have to match the Euclidean solution, i.e the instanton, with the Lorentzian one. The spatial part of the Euclidean de Sitter section is a round 3-sphere, $S^3$. In Euclidean time, they start at zero radius, expand and then contract again, i.e. they form a 4-sphere, $S^4$, with radius $r = \sqrt{3/\Lambda}\,$. The spatial part of the Lorentzian de Sitter universe is again $S^3$, but it expands exponentially in Lorentzian time, forming a 4-hyperboloid. In order to analytically continue the Euclidean to a Lorentzian solution, we join half of the Euclidean $S^4$ with half of the Lorentzian 4-hyperboloid. Only then the matching is smooth from the one sector to the other and the continuity is guaranteed. Therefore, half of the instanton contributes to the solution.

Let us now go through the same procedure but with Schwarzschild-de Sitter spacetime, where the form of $A(r)$ is 
\begin{equation} \label{SdS_metric}
A(r)=1-\frac{2 G_{\text{N}} M}{r}-\frac{\Lambda}{3} r^2\,.
\end{equation}
This spacetime represents a pair of black holes in antipodal points, embedded in a de Sitter space. Thus it is endowed with two horizons; the first one is the event horizon of the black hole $r_{\mathrm{h}}$, while the other one is the cosmological $r_{\mathrm{c}}$ with $r_{\mathrm{h}} < r_{\mathrm{c}}\,$. The topology of SdS spacetime is thus $S^1 \times S^2$, where the radius of $S^2$ varies along $S^1$. The minimum radius is $r_{\mathrm{h}}$ and the maximum $r_{\mathrm{c}}\,$. What we need to do, is to construct a regular instanton, able to be analytically continued to the Lorentzian slice. However, the two horizons  have obviously different surface gravities and, thus, different Hawking temperatures. Hence, we cannot eliminate both of them at once. This is only plausible in the limiting case, where the two horizons coincide (Nariai spacetime) at $\varrho=r_{\mathrm{h}}=r_{\mathrm{c}}=3G_{\text{N}}M=1/\sqrt{\Lambda}\,$. 

In order to see the regularity of the instanton, we transform via the periodic variables
\begin{equation}
\tau = \frac{\psi}{\varepsilon \Lambda} \,,\quad r = \varrho - \varepsilon \cos\xi \,,\quad r_{\mathrm{c}} = \varrho + \varepsilon \,,\quad r_{\mathrm{h}} = \varrho - \varepsilon \,,\,\, \text{where}\,\, \psi \in \left[0,2\pi \right] \,\text{and}\, \xi \in \left[0,\pi \right]\,.
\end{equation}
In the limit $\varepsilon \rightarrow 0$ the two horizons coincide and the form of the metric potential is $A(r) \approx \Lambda \varepsilon^2  \sin^2\xi\,$. Finally, the Nariai instanton becomes
\begin{equation}\label{SdSinstanton}
ds^2 = \frac{1}{\Lambda} \left( d\xi^2 + \sin^2\xi \ d\psi^2 +   d\Omega^2 \right) \,.
\end{equation}
The topology of Nariai spacetime is the product of two round 2-spheres with the same radius, $\varrho = 1/\sqrt{\Lambda}\,$. As before, the matching can be visualized as  joining  half of one of the 2-spheres with half of the Lorentzian 2-hyperboloid. 

The pair creation rate $\Gamma\,$, is now defined from Eq. \eqref{rateformula}. As we have already mentioned in the introduction \ref{intro}, in GR, where the action is the Einstein-Hilbert with a cosmological constant,  de Sitter instanton is found to be $I_{{\mathrm{dS}}} = - 3\pi/\Lambda\,$, while the SdS one is $I_{{\mathrm{SdS}}} = - 2\pi/\Lambda\,$. By taking the half of each and putting them in Eq. \eqref{rateformula} we obtain $\Gamma = e^{-\pi/\Lambda}\,$,
which is determined as the pair creation rate of two SdS black holes in a de Sitter universe. The interpretation of this result is discussed in the introduction.

%Before we proceed to the calculation of the rate in $f(R)-$theories, a few comments are important. It is known that, today the observed value of the cosmological constant is very small, i.e. $\Lambda \sim 10^{-52} m^{-2}$, and thus the production of a pair of black holes, with size $r_{bh} = 1/\sqrt{\Lambda} = 10^{26} \ m$, in today's universe is significantly suppressed \cite{Bousso:1996au,BouHa95}. However, in the early universe and specifically, during inflation, it is thought that the value of $\Lambda \simeq 1$. This effectively means that, even though in the early universe the rate is very large and thus the probability of the creation of two black holes is very high, the size of the black holes to be created is very small. 

\section{Pair creation rate in $f(R)$ models}
\label{fRpcrate}

In this section, we will apply the procedure elaborated in the previous section on the $f(R)-$models that have solutions of the form \eqref{f(R)SdS}. The mathematical form of both de Sitter and Schwarzschild-de Sitter spacetime is the same in GR and in $f(R)$ theories. The only difference is that in $f(R)$, the gravitational coupling and the cosmological constant are not the ones appearing in GR, but some other, effective ones with the influence of the new scalar degree of freedom.

We summarize the results of this section \ref{fRpcrate} in the Table \ref{tab:sum}. In the third column, we show the computed black hole pair creation rate for every model. As we have already discussed, our aim was to set stringent constraints on the theories, in the framework of the NBP. In the last column of Table \ref{tab:sum} we show the obtained constraints for every  $f(R)-$model used in this study.
Further notice that $R_{{\mathrm{dS}}} \propto \Lambda_{{\mathrm{eff}}}$ from Eq.\eqref{lambdaeff}, and so $R_{{\mathrm{dS}}}$ can be regarded as the effective cosmological constant inside the pair creation rate.

\begin{table}[h!]
  \centering
  \begin{tabular}{r||c|c|c}
    \hline
    \hline
    \!\!\!\!\!\! Model  & $f(R)$ \!\,\,\,\,\,\,\,\,\,\,\,  &
    \!\!\!\!\!\! $\Gamma$  &\!\!\!\! constraints\!\!\!\!\!\!  \\
    \hline
    \!\!\! power law  & \!\! $R + \alpha R^n$ & \!\!\! $\exp \left[  - \frac{8 \pi}{R_{{\mathrm{dS}}}} \left( \frac{n-1}{n-2} \right) \right]$  & \!\!\!  $n>2$  \\
    \!\!\! exponential  & \!\! $R - \alpha R_{{\mathrm{dS}}} \left(1-e^{-mR/R_{{\mathrm{dS}}}}\right)$  & \!\!\! $\exp \left[  - \frac{8 \pi}{G_{\text{N}} R_{{\mathrm{dS}}}}\frac{e^m-m-1}{ 2 e^m-m-2}  \right]$  & \!\!\! $m>0$ \\
    \!\!\! $R^2$  & \!\! $\alpha R^2$ & \!\!\! $e^{-8\pi \alpha/G_{\text{N}}}$  & \!\!\! excluded  \\
    \!\!\! Starobinsky  & \!\! $R - \lambda  R_{{\mathrm{dS}}} \left(1 - \left(1+\frac{R^2}{R_{{\mathrm{dS}}}^2}\right)^{-n}\right)$ & \!\!\!  $\exp \left[ -\frac{8 \pi  \left(2^n-n-1\right)}{G_{\text{N}} \left(2^{n+1}-n-2\right) R_{{\mathrm{dS}}}} \right]$ & \!\!\! $n>1$  \\
    \!\!\! Hu-Sawicki  & \!\! $R - \lambda  R_{{\mathrm{dS}}}\frac{\left(\frac{R}{R_{{\mathrm{dS}}}}\right)^{2 m}}{\left(\frac{R}{R_{{\mathrm{dS}}}}\right)^{2 m}+1}$ & \!\!\! $\exp\left[-\frac{8 \pi }{G_{\text{N}} R_{{\mathrm{dS}}}} \frac{m-1}{m-2}\right]$ & \!\!\! $0<m<1$ or $m>2$  \\ 
   \hline
    \hline
  \end{tabular}
  \caption{The overall results of our calculations are presented in this table. The first and second column define the $f(R)-$model used, the third column is the pair creation rate of black holes inside the respected model and in the fourth column the constraints deduced from our analysis.}
  \label{tab:sum}
\end{table} 

\noindent The calculation of every instanton action goes like this. The action \eqref{Iaction} gives
\begin{equation}\label{Iaction1}
I = - \frac{1}{16 \pi G_{\text{N}}} \int d^4x \sqrt{g} \ f(R_{{\mathrm{dS}}}) = - \frac{f(R_{{\mathrm{dS}}})}{16\pi G_{\text{N}}}\int d^4x \sqrt{g}\,,
\end{equation}
since $f(R_{\text{dS}})$ is constant. The integral of Eq. \eqref{Iaction1} for   de Sitter \eqref{dSinstanton} and  SdS \eqref{SdSinstanton} instanton respectively give
\begin{align}
I_{{\mathrm{dS}}} &= - \frac{f(R_{{\mathrm{dS}}})}{16 \pi G_{\text{N}}} \frac{24 \pi ^2 }{\Lambda_{\text{eff}}^2}\,,\\
I_{{\mathrm{SdS}}} &= - \frac{f(R_{{\mathrm{dS}}})}{16 \pi G_{\text{N}}} \frac{16 \pi ^2}{\Lambda ^2 _{\text{eff}}} \,.
\end{align}
By taking half of each and plugging them into Eq. \eqref{rateformula} we obtain
\begin{equation}\label{fRrate}
\Gamma = \exp\left[- \frac{\pi f(R_{{\mathrm{dS}}})}{2 G_{\text{N}} \Lambda_{\mathrm{eff}}^2}\right]=\exp\left[-\frac{2\pi f'(R_{{\mathrm{dS}}})^2}{G_{\text{N}} f(R_{{\mathrm{dS}}})} \right]\,,
\end{equation}
where we used Eq. \eqref{lambdaeff}. This is the main result of the paper and it shows the pair creation rate of two SdS black holes in a de Sitter universe for any $f(R)-$theory of gravity with constant curvature, $R= R_{{\mathrm{dS}}}$. 

As  mentioned in the previous section, in order to avoid ghosts in $f(R)-$theories, one has to assume that $f'(R)>0\,$. Hence, in the rate \eqref{fRrate} we see that $2\pi f'(R_{{\mathrm{dS}}})/G_{\text{N}} >0\,$. If we want to reproduce the same results as in GR, i.e. the rate to be realistic and thus the fraction in the exponent to be negative, meaning that the creation of large black holes in the current universe is suppressed, we have to consider that $f'(R_{{\mathrm{dS}}})/f(R_{{\mathrm{dS}}}) > 0 \,$. That means  $\Lambda _{\text{eff}} >0$ or $R_{{\mathrm{dS}}} > 0$ from Eq. \eqref{lambdaeff}. This is another criterion that we did not mention in the previous section. It is used in order for $f(R)-$models to be cosmologically viable at late times. Therefore, even though the solution of the form  \eqref{f(R)SdS} can have either of SdS type or of SAdS, we see that only in SdS universes the pair creation rate is physically reasonable because we get $\Gamma<1\,$. In other words, if $\Gamma>1$ then de Sitter spacetime is catastrophically unstable to the black hole formation. Moreover, if we consider \textit{a priori} that $\Lambda _{\text{eff}}>0$ and also $f'(R)>0 \,$, then the rate has the same behaviour for any $f(R)-$model; tiny black holes are unsuppressed in an inflationary universe, while large black holes are suppressed at late times.

In the rest of the section, we will apply  Eq. \eqref{fRrate} to specific $f(R)-$models that have been studied extensively in the literature and we will try to constrain their free parameters.

\subsection{The power law $f(R)$}

Let us consider the power law case, where $f(R)  = R + \alpha R^n$ 
%\begin{wraptable}{r}{5cm}
%\begin{center}
%\vspace{-0.6cm}
%\scalebox{0.85}{
%\begin{tabular}{@{}c|c|c}
%\hline  \hline
%a/a& m  & $f(R)$  \\ 
%\hline
%1 & 0  & $R - 2 \Lambda$ \\ 
%2 & 3/2  & $R + R^{3/2}/2\Lambda^{1/2}$ \\ 
%3 & 3  &  $R + R^3/16\Lambda^2$ \\ 
%\hline \hline
%\end{tabular}}
%\end{center}
%\vspace{-0.2cm}
%\caption{Forms of $f(R)$ for different values of $m$. As expected for $m=0$ we recover GR with cosmological constant. The $R^{3/2}$-model is known for reproducing MOND-like behaviour in the weak-field limit and thus reproducing the rotation curves of the galaxies \cite{Capozziello:2017rvz}.}
%\label{table1}
%\end{wraptable}
and $\alpha$ being a constant with the appropriate dimensions, depending on $n$. After substituting the model in the trace equation \eqref{R=R0trace} and solving for $\alpha$ we can see that there can exist black holes of the type \eqref{f(R)SdS}, only for $\alpha = R_{{\mathrm{dS}}}^{1-n}/(n-2)\,$. For $n=2$ this model does not give black hole solutions, since $\alpha$ blows up\footnote{We will consider as a different model the $f(R) = \alpha R^2\,$, which indeed accepts black holes solutions.}.

The rate \eqref{fRrate} becomes
\begin{equation}
\Gamma = \exp \left[  - \frac{8 \pi}{R_{{\mathrm{dS}}}} \left( \frac{n-1}{n-2} \right) \right] \,,
\end{equation}
which has the correct sign only for $n<1$ or $n>2\,$. However, for stability reasons \cite{DeFelice:2010aj}, we have to take into account only $\alpha >0 $ and $n>0$ solutions and, thus, the only acceptable models are those with $n>2\,$.

\subsection{The exponential model}

In this section we study the so-called exponential model, 
\begin{equation}
f(R) = R - \alpha R_{{\mathrm{dS}}} \left(1-e^{-mR/R_{{\mathrm{dS}}}}\right)\,,
\end{equation}
which in the last few years has received a lot of attention in the literature \cite{Chen:2014tdy,Cognola:2007zu,Elizalde:2010ts,Bamba:2010ws,Linder:2009jz,Yang:2010xq}. The constants $\alpha$ and $m$ are two coupling constants with the former being dimensionfull and the latter dimensionless. However, stability of the late-time de Sitter point, together with some more constraints \cite{Bamba:2010ws}, dictate that $\alpha >1\,$. The trace equation \eqref{R=R0trace} is satisfied only for $\alpha \rightarrow \frac{e^m}{2 e^m-m-2}$ (for $m \neq 0$). 

The rate \eqref{fRrate} in this case takes the form
\begin{equation}
\Gamma =\exp \left[  - \frac{8 \pi}{G_{\text{N}} R_{{\mathrm{dS}}}}\frac{e^m-m-1}{ 2 e^m-m-2}  \right]\,,
\end{equation}
and has the right sign only for $m \lesssim -1.59$ or $m>0\,$, from which only the second condition gives the right values for $\alpha\,$.

\subsection{$f(R) = \alpha R^2$}

Another interesting theory is the one described by the $R^2$ Lagrangian density \cite{Kehagias:2015ata}. The coupling constant $\alpha$ scales as a mass term $M^{-2}$ and, thus, has to be positive to avoid tachyonic behaviour. The trace equation \eqref{R=R0trace} is satisfied for any $\alpha\,$, which effectively means that this model has infinite de Sitter points.

Surprisingly enough, the pair creation rate \eqref{fRrate} for this model reads 
\begin{equation}
\Gamma=e^{-8\pi \alpha/G_{\text{N}}}\,.
\end{equation}
As we have already said, $\alpha > 0$ and thus the rate is exponentially decreasing, as wanted. However, we see that in this case, there is no dependence on the effective cosmological constant $\Lambda_{\text{eff}} \sim R_{{\mathrm{dS}}}\,$. This practically means that, since $\alpha$ is considered a constant, the production rate is also a constant\footnote{There is a big discussion in the literature on whether $f(R)-$theories are subject to the so-called \textit{chameleon mechanism} \cite{Khoury:2003rn}. In this case, the mass of the scalaron is considered to be a function of the energy density of its environment. In this way, it can have large mass at short distances, in order to be well-hidden from observations, but at cosmological distances it becomes much lighter, in order to propagate freely. However, since it is still under investigation, for simplicity, we do not consider this case.}. In other words, during inflation, where $\Lambda_{\text{eff}} \sim 1\,$, as well as in current times, where $\Lambda_{\text{eff}} \sim 0\,$, the production rate is the same; which is very unlikely! 
 
\subsection{Starobinsky model}

Here we consider the well known for its inflationary behavior Starobinsky model \cite{Starobinsky:2007hu,Capozziello:2007eu}, which is described by the model
\begin{equation}
f(R) = R - \lambda  R_{{\mathrm{dS}}} \left(1 - \left(1+\frac{R^2}{R_{{\mathrm{dS}}}^2}\right)^{-n}\right)\,,
\end{equation}
with both $\lambda, \,n>0 \,$. In order for the trace equation \eqref{R=R0trace} to be satisfied and, thus, having black hole solutions of the form \eqref{f(R)SdS}, we have to set
$$
\lambda = \frac{2^n}{2^{n+1}-n-2}\,, \quad \,\,\forall \,n \neq 0 , -1\,.
$$
The rate \eqref{fRrate} for this model reads
\begin{equation}
\Gamma = \exp \left[ -\frac{8 \pi  \left(2^n-n-1\right)}{G_{\text{N}} \left(2^{n+1}-n-2\right) R_{{\mathrm{dS}}}} \right]\,,
\end{equation}
which has the correct sign for $n>1\,$, where we have also considered that $n>0\,$. The same bound finds someone if they require that the model should pass laboratory and solar system tests. 
 
\subsection{Hu-Sawicki model}

Last but not least, we study the well-known Hu-Sawicki model \cite{Hu:2007nk,Capozziello:2007eu,Tsujikawa:2007xu}
\begin{equation}
f(R) = R - \lambda  R_{{\mathrm{dS}}}\frac{\left(\frac{R}{R_{{\mathrm{dS}}}}\right)^{2 m}}{\left(\frac{R}{R_{{\mathrm{dS}}}}\right)^{2 m}+1}\,,
\end{equation}
which is very successful in describing the late-time acceleration. The constants $\lambda \,,m$ are considered positive. As in the previous cases, the coupling $\lambda$ has to take the form 
$$
\lambda = - \frac{2}{m-2}\,, \quad \,\,\forall \,m \neq 2\,,
$$
in order for the trace equation to be satisfied and for the model to be able to give SdS-like black holes. 

The rate takes now the form
\begin{equation}
\Gamma = \exp\left[-\frac{8 \pi }{G_{\text{N}} R_{{\mathrm{dS}}}} \frac{m-1}{m-2}\right]\,,
\end{equation}
and $0<m<1$ or $m>2$ are the possible ranges of $m$, in order for the rate to be smaller than unity.

%\subsection{Summary of the pair creation rates}

\section{Conclusions}
\label{conclus}
The inflationary era is the only period when we could expect that quantum pair creation of black holes took place. It is exactly this idea that we use to propose a new criterion for the viability of modified descriptions of the gravitational interactions. Specifically, we claim that \textit{any viable alternative to GR should predict the same behaviour for the black hole pair creation rate;} i.e. exponentially decreasing with the decrease of the effective cosmological constant.

We use the semi-classical instanton method to calculate the probability of a Schwarzschild-de Sitter universe and a de Sitter one, in order to deduce the pair creation of black holes in an expanding universe. We follow this procedure for $f(R)-$theories with constant curvature and calculate the pair creation rate in any of these theories. We claim that, since these theories effectively behave as GR (since all the higher derivative terms in the equations of motion vanish), our results are the same as considering in GR that $\Lambda_{\text{eff}} = \text{constant}$ (in the sense that the solutions are effectively the same). In addition, we study how the rate behaves in well-known $f(R)-$models, again with constant curvature, and set constraints on their free parameters, in order for the rate to have the correct behaviour. We think that even though $R = \text{constant}$ assumption is not generically true during inflation, the results are mathematically interesting, at least as a first approach to the problem. 

As a next step, it would be very interesting to study what happens to specific $f(R)-$models with general black hole solutions (without constant curvature) and how they differ from our results. In addition, the behaviour of (rotating-) black holes with electric and/or magnetic charges, as well as their evolution and stability \cite{Elizalde:1999dw} in modified theories are in our future goals. A final remark could be whether these pair created black holes made it till today. As it was pointed by Bousso \& Hawking \cite{Bousso:1996au}, quantum effects may cause the black hole horizon to be hotter than the cosmological one and thus allow all neutral black holes to evaporate before the end of inflation. In such circumstances, neutrality could be interchanged with a magnetic charge, since this charge cannot be lost and a certain mass is also needed to support such charge. It is our future intention to follow such calculations in $f(R)-$gravity, as well as in other modified theories of gravity.

\acknowledgements
KFD is partly supported by the INFN sezione di Napoli (TEONGRAV). AN is supported by an Alexander von Humboldt
Fellowship. The authors would also like to thank the anonymous referee for her/his comments, which led to the improvement of this article.


\begin{thebibliography}{99}

%\cite{Ade:2015xua}
\bibitem{Ade:2015xua}
  P.~A.~R.~Ade {\it et al.} [Planck Collaboration],
  %``Planck 2015 results. XIII. Cosmological parameters,''
  Astron.\ Astrophys.\  {\bf 594} (2016) A13
%  doi:10.1051/0004-6361/201525830
%  [arXiv:1502.01589 [astro-ph.CO]].
  %%CITATION = doi:10.1051/0004-6361/201525830;%%
  %4339 citations counted in INSPIRE as of 08 Nov 2017


\bibitem{CaH74}
  B.~J.~Carr and S.~W.~Hawking,
  %``Black Holes in the Early Universe,''
  Mon.\ Not.\ Roy.\ Astron.\ Soc. {\bf 168} (1974) 399-415
%  https://doi.org/10.1093/mnras/168.2.399


\bibitem{Car04}
  B.~J.~Carr,
  %``Primordial black holes: Recent developments,''
  eConf {\bf C041213} (2004) 0204
  % reportNumber   = "TSRA-2004-0204"%
  % astro-ph/0504034%
%%CITATION = ASTRO-PH/0504034;%%

	
	
\bibitem{RSK01}
 S.~G.~Rubin, A.~S.~Sakharov and  M.~Yu.~Khlopov,                       
  %``The Formation of primary galactic nuclei during phase transitions in the early universe,''
 J.\ Exp.\ Theor.\ Phys. {\bf 92} (2001) 921-929
%  doi:10.1134/1.1385631
%  hep-ph/0106187
  %%CITATION = HEP-PH/0106187;%%

	

\bibitem{Khl10}
   M.~Yu.~Khlopov,                       
  %``Primordial Black Holes,''
  Res.\ Astron.\ Astrophys. {\bf 10}  (2010) 495-528
%  doi:10.1088/1674-4527/10/6/001
%  0801.0116.
  %%CITATION = ARXIV:0801.0116;;%%



\bibitem{KRS05}
 M.~Yu.~Khlopov, S.~G.~Rubin and A.~S.~Sakharov,                         
  %``Primordial structure of massive black hole clusters,''
 Astropart.\ Phys. {\bf 23} (2005) 265
%  doi:10.1016/j.astropartphys.2004.12.002
%  astro-ph/0401532
  %%CITATION = ASTRO-PH/0401532%%


%\cite{Mann:1995vb}
\bibitem{Mann:1995vb}
  R.~B.~Mann and S.~F.~Ross,
  %``Cosmological production of charged black hole pairs,''
  Phys.\ Rev.\ D {\bf 52} (1995) 2254
%  doi:10.1103/PhysRevD.52.2254
%  [gr-qc/9504015].
  %%CITATION = doi:10.1103/PhysRevD.52.2254;%%
  %135 citations counted in INSPIRE as of 08 Nov 2017	

%\cite{Dowker:1993bt}
\bibitem{Dowker:1993bt}
  F.~Dowker, J.~P.~Gauntlett, D.~A.~Kastor and J.~H.~Traschen,
  %``Pair creation of dilaton black holes,''
  Phys.\ Rev.\ D {\bf 49} (1994) 2909
%  doi:10.1103/PhysRevD.49.2909
%  [hep-th/9309075].
  %%CITATION = doi:10.1103/PhysRevD.49.2909;%%
  %216 citations counted in INSPIRE as of 08 Nov 2017
  
  %\cite{Bousso:1996au}
\bibitem{Bousso:1996au}
  R.~Bousso and S.~W.~Hawking,
  %``Pair creation of black holes during inflation,''
  Phys.\ Rev.\ D {\bf 54} (1996) 6312
%  doi:10.1103/PhysRevD.54.6312
%  [gr-qc/9606052].
  %%CITATION = doi:10.1103/PhysRevD.54.6312;%%
  %161 citations counted in INSPIRE as of 08 Nov 2017
  
  %\cite{Garfinkle:1993xk}
\bibitem{Garfinkle:1993xk}
  D.~Garfinkle, S.~B.~Giddings and A.~Strominger,
  %``Entropy in black hole pair production,''
  Phys.\ Rev.\ D {\bf 49} (1994) 958
%  doi:10.1103/PhysRevD.49.958
%  [gr-qc/9306023].
  %%CITATION = doi:10.1103/PhysRevD.49.958;%%
  %102 citations counted in INSPIRE as of 08 Nov 2017
  
  %\cite{Dowker:1994up}
\bibitem{Dowker:1994up}
  F.~Dowker, J.~P.~Gauntlett, S.~B.~Giddings and G.~T.~Horowitz,
  %``On pair creation of extremal black holes and Kaluza-Klein monopoles,''
  Phys.\ Rev.\ D {\bf 50} (1994) 2662
%  doi:10.1103/PhysRevD.50.2662
%  [hep-th/9312172].
  %%CITATION = doi:10.1103/PhysRevD.50.2662;%%
  %172 citations counted in INSPIRE as of 08 Nov 2017
  
  %\cite{Hawking:1994ii}
\bibitem{Hawking:1994ii}
  S.~W.~Hawking, G.~T.~Horowitz and S.~F.~Ross,
  %``Entropy, Area, and black hole pairs,''
  Phys.\ Rev.\ D {\bf 51} (1995) 4302
%  doi:10.1103/PhysRevD.51.4302
%  [gr-qc/9409013].
  %%CITATION = doi:10.1103/PhysRevD.51.4302;%%
  %345 citations counted in INSPIRE as of 08 Nov 2017
  
  %\cite{Hawking:1995ap}
\bibitem{Hawking:1995ap}
  S.~W.~Hawking and S.~F.~Ross,
  %``Duality between electric and magnetic black holes,''
  Phys.\ Rev.\ D {\bf 52} (1995) 5865
%  doi:10.1103/PhysRevD.52.5865
%  [hep-th/9504019].
  %%CITATION = doi:10.1103/PhysRevD.52.5865;%%
  %163 citations counted in INSPIRE as of 08 Nov 2017

%\cite{Caldwell:1996pt}
\bibitem{Caldwell:1996pt}
  R.~R.~Caldwell, H.~A.~Chamblin and G.~W.~Gibbons,
  %``Pair creation of black holes by domain walls,''
  Phys.\ Rev.\ D {\bf 53} (1996) 7103
%  doi:10.1103/PhysRevD.53.7103
%  [hep-th/9602126].
  %%CITATION = doi:10.1103/PhysRevD.53.7103;%%
  %56 citations counted in INSPIRE as of 08 Nov 2017
  
\bibitem{MaN11}
R.~B. Mann, P.~Nicolini, 
%\enquote{{Cosmological production of noncommutative black holes},} 
Phys. \ Rev. \ D {\bf 84}, \ 064014 (2011).

    %\cite{HarHawk83}
\bibitem{HarHawk83}
  J. B. Hartle, S. W. Hawking,
  %``The wave function of the universe,''
  Phys.\ Rev.\ D {\bf 28},\ 2960\ (1983) 
%  doi:https://doi.org/10.1103/PhysRevD.28.2960


%\cite{Capozziello:2011et}
\bibitem{Capozziello:2011et}
  S.~Capozziello and M.~De Laurentis,
  %``Extended Theories of Gravity,''
  Phys.\ Rept.\  {\bf 509} (2011) 167
%  doi:10.1016/j.physrep.2011.09.003
%  [arXiv:1108.6266 [gr-qc]].
  %%CITATION = doi:10.1016/j.physrep.2011.09.003;%%
  %870 citations counted in INSPIRE as of 08 Nov 2017

%\cite{Clifton:2011jh}
\bibitem{Clifton:2011jh}
  T.~Clifton, P.~G.~Ferreira, A.~Padilla and C.~Skordis,
  %``Modified Gravity and Cosmology,''
  Phys.\ Rept.\  {\bf 513} (2012) 1
%  doi:10.1016/j.physrep.2012.01.001
%  [arXiv:1106.2476 [astro-ph.CO]].
  %%CITATION = doi:10.1016/j.physrep.2012.01.001;%%
  %1321 citations counted in INSPIRE as of 08 Nov 2017

%\cite{Capozziello:2009nq}
\bibitem{Capozziello:2009nq}
  S.~Capozziello, M.~De Laurentis and V.~Faraoni,
  %``A Bird's eye view of f(R)-gravity,''
  Open Astron.\ J.\  {\bf 3} (2010) 49
%  doi:10.2174/1874381101003010049, 10.2174/1874381101003020049
%  [arXiv:0909.4672 [gr-qc]].
  %%CITATION = doi:10.2174/1874381101003010049, 10.2174/1874381101003020049;%%
  %160 citations counted in INSPIRE as of 08 Nov 2017

%\cite{Nojiri:2010wj}
\bibitem{Nojiri:2010wj}
  S.~Nojiri and S.~D.~Odintsov,
  %``Unified cosmic history in modified gravity: from F(R) theory to Lorentz non-invariant models,''
  Phys.\ Rept.\  {\bf 505} (2011) 59
%  doi:10.1016/j.physrep.2011.04.001
%  [arXiv:1011.0544 [gr-qc]].
  %%CITATION = doi:10.1016/j.physrep.2011.04.001;%%
  %1673 citations counted in INSPIRE as of 17 Apr 2018

%\cite{Nojiri:2006ri}
\bibitem{Nojiri:2006ri}
  S.~Nojiri and S.~D.~Odintsov,
  %``Introduction to modified gravity and gravitational alternative for dark energy,''
  eConf C {\bf 0602061} (2006) 06
   [Int.\ J.\ Geom.\ Meth.\ Mod.\ Phys.\  {\bf 4} (2007) 115]
%  doi:10.1142/S0219887807001928
%  [hep-th/0601213].
  %%CITATION = doi:10.1142/S0219887807001928;%%
  %1818 citations counted in INSPIRE as of 17 Apr 2018



\bibitem{BouHa95}
  R. Bousso, S. W. Hawking,
  %``The Probability for Primordial Black Holes,''
  Phys.\ Rev.\ D52\ (1995)\ 5659-5664
 % doi:https:10.1103/PhysRevD.52.5659
  %[arXiv:gr-qc/9506047].

%\cite{Bousso:1996wz}
\bibitem{Bousso:1996wz}
  R.~Bousso and S.~W.~Hawking,
  %``Pair creation and evolution of black holes in inflation,''
  Helv.\ Phys.\ Acta {\bf 69} (1996) 261
%  [gr-qc/9608008].
  %%CITATION = GR-QC/9608008;%%
  %5 citations counted in INSPIRE as of 24 Dec 2017

%\cite{Vilenkin:1986cy}
\bibitem{Vilenkin:1986cy}
  A.~Vilenkin,
  %``Boundary Conditions in Quantum Cosmology,''
  Phys.\ Rev.\ D {\bf 33} (1986) 3560.
%  doi:10.1103/PhysRevD.33.3560
  %%CITATION = doi:10.1103/PhysRevD.33.3560;%%
  %366 citations counted in INSPIRE as of 04 Dec 2017

%\cite{DeFelice:2010aj}
\bibitem{DeFelice:2010aj}
  A.~De Felice and S.~Tsujikawa,
  %``f(R) theories,''
  Living Rev.\ Rel.\  {\bf 13} (2010) 3
 % doi:10.12942/lrr-2010-3
 % [arXiv:1002.4928 [gr-qc]].
  %%CITATION = doi:10.12942/lrr-2010-3;%%
  %1410 citations counted in INSPIRE as of 01 Dec 2017

%\cite{Sotiriou:2008rp}
\bibitem{Sotiriou:2008rp}
  T.~P.~Sotiriou and V.~Faraoni,
  %``f(R) Theories Of Gravity,''
  Rev.\ Mod.\ Phys.\  {\bf 82} (2010) 451
%  doi:10.1103/RevModPhys.82.451
%  [arXiv:0805.1726 [gr-qc]].
  %%CITATION = doi:10.1103/RevModPhys.82.451;%%
  %1910 citations counted in INSPIRE as of 01 Dec 2017
  
%\cite{delaCruzDombriz:2009et}
\bibitem{delaCruzDombriz:2009et}
  A.~de la Cruz-Dombriz, A.~Dobado and A.~L.~Maroto,
  %``Black Holes in f(R) theories,''
  Phys.\ Rev.\ D {\bf 80} (2009) 124011
 %  Erratum: [Phys.\ Rev.\ D {\bf 83} (2011) 029903]
%  doi:10.1103/PhysRevD.83.029903, 10.1103/PhysRevD.80.124011
%  [arXiv:0907.3872 [gr-qc]].
  %%CITATION = doi:10.1103/PhysRevD.83.029903, 10.1103/PhysRevD.80.124011;%%
  %120 citations counted in INSPIRE as of 08 Nov 2017  
  
%\cite{Capozziello:2011wg}
\bibitem{Capozziello:2011wg}
  S.~Capozziello and D.~Saez-Gomez,
  %``Scalar-tensor representation of $f(R)$ gravity and Birkhoff's theorem,''
  Annalen Phys.\  {\bf 524} (2012) 279
%  doi:10.1002/andp.201100244
%  [arXiv:1107.0948 [gr-qc]].
  %%CITATION = doi:10.1002/andp.201100244;%%
  %18 citations counted in INSPIRE as of 24 Nov 2017
  
  %\cite{Nzioki:2013lca}
\bibitem{Nzioki:2013lca}
  A.~M.~Nzioki, R.~Goswami and P.~K.~S.~Dunsby,
  %``Jebsen-Birkhoff theorem and its stability in f(R) gravity,''
  Phys.\ Rev.\ D {\bf 89} (2014) no.6,  064050
%  doi:10.1103/PhysRevD.89.064050
%  [arXiv:1312.6790 [gr-qc]].
  %%CITATION = doi:10.1103/PhysRevD.89.064050;%%
  %18 citations counted in INSPIRE as of 24 Nov 2017

\bibitem{Kiefer13}
  C.~Kiefer,
  %``Conceptual Problems in Quantum Gravity and Quantum Cosmology,''
  ISRN Math.\ Phys.\ (2013) 509316 
%  doi:10.1155/2013/509316
%  [arXiv:1401.3578 [gr-qc]].
  


\bibitem{GibHawk13}
  G.W.~Gibbons,   S.W.~Hawking,
  %``Euclidean Quantum Gravity,''
  World\ Scientific\ Publishing\ Company\ (May 1, 1993) 
  


%\cite{Chen:2014tdy}
\bibitem{Chen:2014tdy}
  Y.~Chen, C.~Q.~Geng, C.~C.~Lee, L.~W.~Luo and Z.~H.~Zhu,
  %``Constraints on the exponential $f(R)$ model from latest Hubble parameter measurements,''
  Phys.\ Rev.\ D {\bf 91} (2015) no.4,  044019
%  doi:10.1103/PhysRevD.91.044019
 % [arXiv:1407.4303 [astro-ph.CO]].
  %%CITATION = doi:10.1103/PhysRevD.91.044019;%%
  %6 citations counted in INSPIRE as of 07 Nov 2017

%\cite{Bamba:2010ws}
\bibitem{Bamba:2010ws}
  K.~Bamba, C.~Q.~Geng and C.~C.~Lee,
  %``Cosmological evolution in exponential gravity,''
  JCAP {\bf 1008} (2010) 021
%  doi:10.1088/1475-7516/2010/08/021
%  [arXiv:1005.4574 [astro-ph.CO]].
  %%CITATION = doi:10.1088/1475-7516/2010/08/021;%%
  %89 citations counted in INSPIRE as of 03 Dec 2017

%\cite{Cognola:2007zu}
\bibitem{Cognola:2007zu}
  G.~Cognola, E.~Elizalde, S.~Nojiri, S.~D.~Odintsov, L.~Sebastiani and S.~Zerbini,
  %``A Class of viable modified f(R) gravities describing inflation and the onset of accelerated expansion,''
  Phys.\ Rev.\ D {\bf 77} (2008) 046009
%  doi:10.1103/PhysRevD.77.046009
 % [arXiv:0712.4017 [hep-th]].
  %%CITATION = doi:10.1103/PhysRevD.77.046009;%%
  %392 citations counted in INSPIRE as of 07 Nov 2017

%\cite{Elizalde:2010ts}
\bibitem{Elizalde:2010ts}
  E.~Elizalde, S.~Nojiri, S.~D.~Odintsov, L.~Sebastiani and S.~Zerbini,
  %``Non-singular exponential gravity: a simple theory for early- and late-time accelerated expansion,''
  Phys.\ Rev.\ D {\bf 83} (2011) 086006
%  doi:10.1103/PhysRevD.83.086006
 % [arXiv:1012.2280 [hep-th]].
  %%CITATION = doi:10.1103/PhysRevD.83.086006;%%
  %106 citations counted in INSPIRE as of 07 Nov 2017

%\cite{Linder:2009jz}
\bibitem{Linder:2009jz}
  E.~V.~Linder,
  %``Exponential Gravity,''
  Phys.\ Rev.\ D {\bf 80} (2009) 123528
%  doi:10.1103/PhysRevD.80.123528
%  [arXiv:0905.2962 [astro-ph.CO]].
  %%CITATION = doi:10.1103/PhysRevD.80.123528;%%
  %135 citations counted in INSPIRE as of 07 Nov 2017


%\cite{Yang:2010xq}
\bibitem{Yang:2010xq}
  L.~Yang, C.~C.~Lee, L.~W.~Luo and C.~Q.~Geng,
  %``Observational Constraints on Exponential Gravity,''
  Phys.\ Rev.\ D {\bf 82} (2010) 103515
%  doi:10.1103/PhysRevD.82.103515
%  [arXiv:1010.2058 [astro-ph.CO]].
  %%CITATION = doi:10.1103/PhysRevD.82.103515;%%
  %33 citations counted in INSPIRE as of 07 Nov 2017
  
  %\cite{Kehagias:2015ata}
\bibitem{Kehagias:2015ata}
  A.~Kehagias, C.~Kounnas, D.~Lüst and A.~Riotto,
  %``Black hole solutions in $R^{2}$ gravity,''
  JHEP {\bf 1505} (2015) 143
%  doi:10.1007/JHEP05(2015)143
%  [arXiv:1502.04192 [hep-th]].
  %%CITATION = doi:10.1007/JHEP05(2015)143;%%
  %27 citations counted in INSPIRE as of 03 Dec 2017

%\cite{Khoury:2003rn}
\bibitem{Khoury:2003rn}
  J.~Khoury and A.~Weltman,
  %``Chameleon cosmology,''
  Phys.\ Rev.\ D {\bf 69} (2004) 044026
%  doi:10.1103/PhysRevD.69.044026
%  [astro-ph/0309411].
  %%CITATION = doi:10.1103/PhysRevD.69.044026;%%
  %940 citations counted in INSPIRE as of 30 Nov 2017

%\cite{Starobinsky:2007hu}
\bibitem{Starobinsky:2007hu}
  A.~A.~Starobinsky,
  %``Disappearing cosmological constant in f(R) gravity,''
  JETP Lett.\  {\bf 86} (2007) 157
%  doi:10.1134/S0021364007150027
%  [arXiv:0706.2041 [astro-ph]].
  %%CITATION = doi:10.1134/S0021364007150027;%%
  %675 citations counted in INSPIRE as of 03 Dec 2017
  
  %\cite{Capozziello:2007eu}
\bibitem{Capozziello:2007eu}
  S.~Capozziello and S.~Tsujikawa,
  %``Solar system and equivalence principle constraints on f(R) gravity by chameleon approach,''
  Phys.\ Rev.\ D {\bf 77} (2008) 107501
%  doi:10.1103/PhysRevD.77.107501
%  [arXiv:0712.2268 [gr-qc]].
  %%CITATION = doi:10.1103/PhysRevD.77.107501;%%
  %188 citations counted in INSPIRE as of 03 Dec 2017
  
  %\cite{Tsujikawa:2007xu}
\bibitem{Tsujikawa:2007xu}
  S.~Tsujikawa,
  %``Observational signatures of f(R) dark energy models that satisfy cosmological and local gravity constraints,''
  Phys.\ Rev.\ D {\bf 77} (2008) 023507
%  doi:10.1103/PhysRevD.77.023507
%  [arXiv:0709.1391 [astro-ph]].
  %%CITATION = doi:10.1103/PhysRevD.77.023507;%%
  %275 citations counted in INSPIRE as of 03 Dec 2017 


%\cite{Hu:2007nk}
\bibitem{Hu:2007nk}
  W.~Hu and I.~Sawicki,
  %``Models of f(R) Cosmic Acceleration that Evade Solar-System Tests,''
  Phys.\ Rev.\ D {\bf 76} (2007) 064004
%  doi:10.1103/PhysRevD.76.064004
%  [arXiv:0705.1158 [astro-ph]].
  %%CITATION = doi:10.1103/PhysRevD.76.064004;%%
  %1000 citations counted in INSPIRE as of 03 Dec 2017  

%\cite{Elizalde:1999dw}
\bibitem{Elizalde:1999dw}
  E.~Elizalde, S.~Nojiri and S.~D.~Odintsov,
  %``Possible quantum instability of primordial black holes,''
  Phys.\ Rev.\ D {\bf 59} (1999) 061501
  %doi:10.1103/PhysRevD.59.061501
  %[hep-th/9901026].
  %%CITATION = doi:10.1103/PhysRevD.59.061501;%%
  %32 citations counted in INSPIRE as of 17 Apr 2018
 

\end{thebibliography}
\end{document}